\documentclass[acus]{JAC2011}

\usepackage{graphicx}
\usepackage{booktabs}

\begin{document}
\title{A Feasibility Study of an $e^+e^-$ Ring Collider for Higgs Factory\thanks{Work supported by the Department of Energy under
Contract Number: DE-AC02-76SF00515.}}

\author{Yunhai Cai, Alex Chao, Yuri Nosochkov, Uli Wienands, Frank Zimmermann\thanks{on leave from CERN, Geneva, Switzerland} \\
SLAC National Accelerator Laboratory, Menlo Park, CA 94025, USA}

\maketitle

\begin{abstract}

A ring-based Higgs factory with  a center-of-mass energy of 240 GeV residing in the existing LEP tunnel is studied at a level of concrete lattice. We found that low-emittance lattice is essential to mitigate the effect of beamstrahlung on the beam lifetime. To achieve a luminosity of $1.0\times10^{34}{\rm cm}^{-2}{\rm s}^{-1}$, we simplified final focusing system and improved its momentum aperture by a factor of 5 to 2.5\%.

\end{abstract}

\section{Introduction}

Since the discovery of a ``Higgs like" particle at LHC on July 4, 2012, the recent results for ATLAS and CMS have shown that the discovered particle resembles more and more the Higgs boson in the standard model of elementary particles. Because of this remarkable discovery, it will become increasingly important to precisely measure the property of the particle that gives the mass to all others.

Obviously, an $e^+e^-$ storage ring can be an efficient option to provide such a collider for precision measurement because of the low mass of the Higgs boson. Using the production channel  of $e^+e^-\rightarrow HZ$, the minimum beam energy required for such a collider is about 120 GeV, which is approximately 15\% higher than the energy reached 15 years ago at LEP2. After more than a decade, can we design and build this simplest Higgs factory (HF) in the same tunnel?

\section{Luminosity}

In a collider, aside from its energy, its luminosity is the most important design parameter. For Gaussian beams, we can write the bunch luminosity as
\begin{equation}
{\cal L}_b = f_{0} {N_b^2\over 4\pi\sigma_x\sigma_y} R_h,
\end{equation}
where $f_{0}$ is the revolution frequency, $N_b$ the bunch population, $\sigma_{x,y}$ transverse beam sizes, and $R_h$ is a factor of geometrical reduction due to a finite bunch length $\sigma_z$ and is given by
\begin{equation}
R_h = \sqrt{2\over\pi}a e^{a^2} K_0(a^2),
\end{equation}
$a=\beta_y^*/(\sqrt{2}\sigma_z$), $\beta_y^*$ is the vertical beta function at the interaction point (IP), and $K_0$ the modified Bessel function. In order to avoid $R_h$ becoming too small, $\sigma_z \approx \beta_y^*$. Obviously, for a number of $n_b$ bunches, the total luminosity is ${\cal L}=n_b {\cal L}_b$.

In general, the beam sizes in the luminosity formula are dynamic variables. They are subject to the influence of the electromagnetic interaction during collision. To take this collision effect into account, we introduce the beam-beam parameter as
\begin{equation}
\xi_y = {r_e N_b \beta_y^*\over 2\pi\gamma\sigma_y(\sigma_x+\sigma_y)},
\end{equation}
where $\gamma$ is the Lorentz factor and $r_e$ the classical electron radius. Using this formula of $\xi_y$, we can rewrite the luminosity as
\begin{equation}
{\cal L}={cI\gamma\xi_y\over 2 r_e^2 I_A \beta_y^*} R_h,
\end{equation}
where $I$ is the beam current and $I_A=ec/r_e\approx 17045$ A, the Alfven current. Since $\xi_y$ is limited below 0.1 in most colliders, this formula is often used for estimating an upper bound of the luminosity.

\begin{table}[h]
\caption{Main parameters of a Higgs factory based on the LEP tunnel. The LEP2 parameters are tabulated for comparison.}
\begin{center}
\begin{tabular}{lll}
\hline
\hline
Parameter                              & LEP2                        & HF                        \\
\hline
Beam energy, $E_0$ [GeV]               & $104.5$                     & $120.0$                   \\
Circumference, $C$ [km]                & $26.7$                      & $26.7$                    \\
Beam current, $I$ [mA]                 & $4$                         & $7.2$                     \\
SR power, $P_{SR}$ [MW]                & $11$                        & $50$                      \\
Beta function at IP, $\beta_y^*$ [mm]  & $50$                        & $1$                       \\
Bunch length, $\sigma_z$ [mm]          & $16.1$                      & $1.5$                     \\
Hourglass factor, $R_h$                & $0.98$                      & $0.76$                    \\
Beam-beam parameter, $\xi_y$           & $0.07$                      & $0.07$                    \\
Luminosity, $\cal L$ [$10^{34}{\rm cm}^{-2}{\rm s}^{-1}$]               & $0.0125$                       & $1.01$                     \\
\hline
\hline
\end{tabular}
\end{center}
\label{tab:parameter}
\end{table}

In Table \ref{tab:parameter}, we tabulated a set of consistent parameters for a Higgs Factory that can reside in the LEP tunnel. Note that our parameters here are similar to those proposed in the LEP3~\cite{LEP3} design except for the bunch length $\sigma_z$. This difference is due to a much smaller momentum compaction factor $\alpha_p$ in a low emittance lattice as we will show later. In contrast to the other factories, the beam current is severely limited by the power of synchrotron radiation at very high energy.

\section{Synchrotron radiation}

When an electron is in circular motion with a bending radius $\rho$, its energy loss per turn to synchrotron radiation is given by
\begin{equation}
U_0 = {4\pi r_e mc^2\gamma^4\over3\rho}.
\end{equation}
This loss has to be compensated by an RF system. The required RF power per ring is
\begin{equation}
P_{SR}=U_0 I/e.
\end{equation}
For the beam parameters in Table \ref{tab:parameter} and $\rho=2.6$ km, we have $U_0=6.97$ GeV, which means that electron loses about 6\% of its energy every turn. Assuming $P_{SR}$ has to be less than 50 MW, the beam current is limited to 7.2 mA in the ring. Applying the expression of $P_{SR}$ to the luminosity formula, we obtain
\begin{equation}
{\cal L}={3c\xi_y\rho P_{SR}\over 8\pi r_e^3 \gamma^3 \beta_y^* P_A} R_h,
\end{equation}
where $P_A = mc^2 I_A/e \approx 8.7$ GW. This scaling property of luminosity in $e^+e^-$ colliders at extremely high energy was first given by Richter~\cite{richter}.

For a Higgs factory with beam energy larger than 120 GeV, its beam current will be severely capped by the electrical power consumed by the RF system and therefore a smaller $\beta_y^*$ seems the only convenient option to reach a required luminosity.

\section{Beamstrahlung Effects}

Another important aspect of very high energy colliding beams is the emission of photons during collision and sometimes a pair of $e^+$ and $e^-$ can be created in the process. In general, this phenomenon is well known and called the beamstrahlung effect. Recently, Telnov found~\cite{telnov} that the most limiting effects to Higgs factory is an event when a high-energy photon is emitted by an electron in the beamstrahlung process. The electron energy loss can be so large that it falls outside of the momentum aperture $\eta$ in the colliding ring. For a typical Higgs factory, is was suggested that the following,
\begin{equation}
{N_b\over\sigma_x\sigma_z} < {0.1\eta\alpha\over 3\gamma r_e^2},
\end{equation}
has to be satisfied to achieve 30 minutes of beam lifetime. Here $\alpha \approx 1/137$ is the fine structure constant. If we introduce aspect ratios of beta functions at the IP and emittances in the ring, namely $\kappa_\beta=\beta_y^*/\beta_x^*$ and $\kappa_e=\epsilon_y/\epsilon_x$, this criteria can be rewritten as
\begin{equation}
{N_b\over\sqrt{\epsilon_x}} < {0.1\eta\alpha\sigma_z\over3\gamma r_e^2}\sqrt{\beta_y^*\over\kappa_\beta}
\end{equation}.

On the other hand, to achieve the beam-beam parameter $\xi_y$, we need
\begin{equation}
{N_b\over\epsilon_x} = {2\pi\gamma\xi_y\over r_e} \sqrt{\kappa_e\over\kappa_\beta}.
\label{eqn:xi}
\end{equation}
Combining this equation with the lifetime condition, we have
\begin{equation}
\epsilon_x < {\beta_y^*\over\kappa_e}\left({0.1\eta\alpha\sigma_z\over 6\pi\gamma^2\xi_y r_e}\right)^2.
\label{eqn:lifetime}
\end{equation}
Since the quantities like $\xi_y$, $\beta_y^*$, and $\sigma_z$ are largely determined by the required luminosity and $\gamma$ by the particle to be studied, this inequality specifies a low-emittance lattice that is required to achieve 30 minutes of beam lifetime. Normally, the natural emittance scales as $\gamma^2$. Here it requires a scaling of $\gamma^{-4}$, indicating another difficulty to design a factory with much higher energy beyond 120 GeV.

\begin{table}[h]
\caption{Additional parameters selected to mitigate the beamstrahlung effects and reach 30 minutes in beamstrahlung beam lifetime.}
\begin{center}
\begin{tabular}{lll}
\hline
\hline
Parameter                                      & LEP2                     & HF                        \\
\hline
Beam energy, $E_0$ [GeV]                       & $104.5$                  & $120.0$                   \\
Circumference, $C$ [km]                        & $26.7$                   & $26.7$                    \\
Horizontal emittance, $\epsilon_x$ [nm]        & $48$                     & $4.3$                     \\
Vertical emittance, $\epsilon_y$ [nm]          & $0.25$                   & $0.0108$                  \\
Momentum acceptance, $\eta$ [\%]               & $1.0$                    & $2.0$                     \\
Momentum compaction, $\alpha_p$ [$10^{-5}$]    & $18.5$                   & $2.4$                     \\
\hline
\hline
\end{tabular}
\end{center}
\label{tab:para_lifetime}
\end{table}

As shown in Table \ref{tab:para_lifetime}, we need design a lattice with much smaller emittance than the one in LEP2 to mitigate the beamstrahlung effect. In general, a lower emittance requires more magnets in a ring.

\section{Arc lattice}

\begin{figure}[ht]
\includegraphics{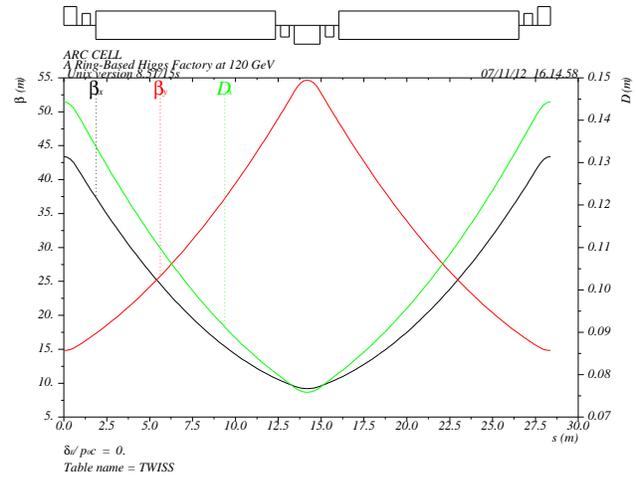}
 \vspace{6.5cm}
 \caption{\it Lattice functions in a $90^0/60^0$ FODO cell. The cell length is more than a factor of two shorter than the one in LEP2.}
\label{fig:cell}
\end{figure}

An optical design of a simple cell is illustrated in Fig.~\ref{fig:cell}. Every twelve such cells make a quasi 4th-order achromat, meaning that the only non-vanishing resonance up to 4th-order is $4\nu_x$. In our design, each arc consists of eight achromats and ends with dispersion suppressors. To fit in the LEP tunnel, we have eight arcs and eight straight sections to complete a ring with parameters shown in Table \ref{tab:para_lifetime}. This ring without a final focusing has a momentum acceptance $\eta > 4\%$.

\section{Final focusing}

Note that the beam lifetime condition in Eq.~(\ref{eqn:lifetime}) does not depend on $\kappa_\beta$. Therefore, according to Eq.~(\ref{eqn:xi}), $\kappa_\beta$ (or $\beta_x^*$) can be used to adjust the bunch population $N_b$ or equivalently the number of bunches $n_b$ when the total current is limited by the electrical power. Here we would like to choose a large $\beta_x^*$, leading to a smaller $n_b$. Our choice of the parameters in the interaction region are tabulated in Table \ref{tab:para_ffs}.

\begin{table}[h]
\caption{Other parameters determined by a specific design of final focusing system.}
\begin{center}
\begin{tabular}{lll}
\hline
\hline
Parameter                                      & LEP2                     & HF                        \\
\hline
Beam energy, $E_0$ [GeV]                       & $104.5$                  & $120.0$                   \\
Circumference, $C$ [km]                        & $26.7$                   & $26.7$                    \\
$\beta_x^*$ [mm]                               & $1500$                   & $100$                     \\
$\beta_y^*$ [mm]                               & $50$                     & $1$                       \\
Bunch population, $N_b$ [$10^{10}$]            & $57.5$                   & $8.0$                     \\
Number of bunches, $n_b$                       & $4$                      & $50$                      \\
\hline
\hline
\end{tabular}
\end{center}
\label{tab:para_ffs}
\end{table}

\begin{figure}[ht]
\includegraphics{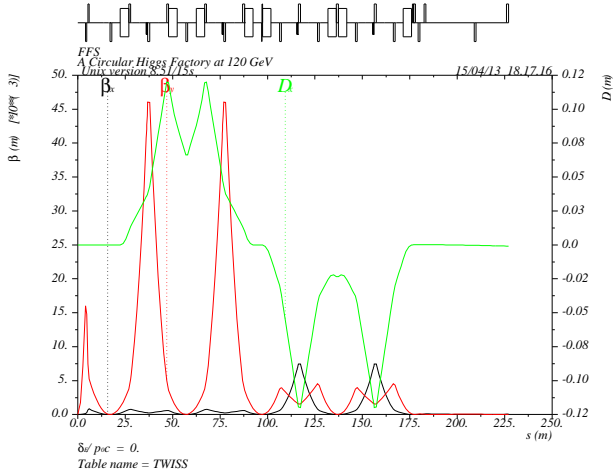}
 \vspace{6.5cm}
 \caption{\it Lattice functions in a final focusing system. It provides a 2.5\% momentum aperture in the circular Higgs factory.}
\label{fig:ffs}
\end{figure}

It is always challenging to design a final focusing system in a collider. In a Higgs factory, it becomes more so because of a smaller $\beta_y^*=1$ mm and a longer distance $L^*=4$ meter: between the IP and the first focusing quadrupole. Here we adopt an optics similar to the design of NLC with two pairs of non-interleaved sextupoles for local chromatic compensation. The optics of a half of the interaction region (IR) is shown in Fig.~\ref{fig:ffs}. The entire IR fits in a 500-meter long straight section.

\begin{figure}[ht]
\includegraphics{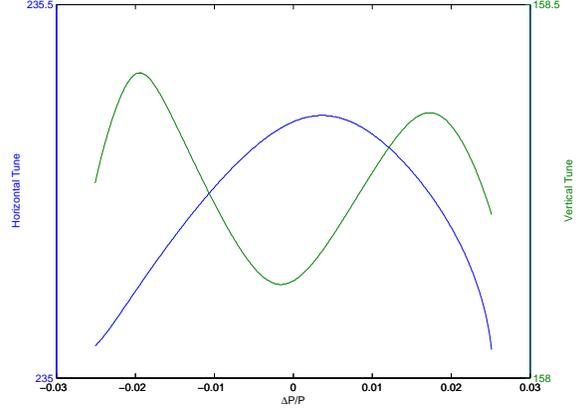}
 \vspace{6cm}
 \caption{\it Betatron tunes of the whole ring as a function of relative momentum.}
\label{fig:bandwidth}
\end{figure}

In addition to the two pairs in the IR, we used two global families of sextupoles in the arcs for the optimization of chromaticity. A typical momentum bandwidth for such a system is about $\pm0.5\%$. As shown in Fig.~\ref{fig:bandwidth}, we enlarged the bandwidth by a factor of five to $\pm2.5\%$. A major improvement came after we significantly reduced the number of IR quadrupoles, which are the ultimate sources of the nonlinear chromatic abberations. At the edges of the momentum bandwidth, we saw steep walls rising in the beta functions. Obviously, we have achieved a good beam lifetime since 2\% of momentum acceptance is required to achieve 30 minutes of beam lifetime as tabulated in Table \ref{tab:para_lifetime}.

\section{Conclusion}

In this paper, the impact on design due to beamstrahlung is analyzed. We found a formula for natural emittance that is necessary to achieve an adequate beam lifetime. As a result, a systematic design procedure is also given in this paper.

Our analysis shows that circular Higgs factory requires not only a final focusing system with an ultra-low beta but also a low-emittance lattice at very high energy. Moreover, based on this preliminary study, we believe that it is feasible to design and to build a circular Higgs factory that has a luminosity of $1.0\times10^{34}{\rm cm}^{-2}{\rm s}^{-1}$ and an adequate beam lifetime in the existing LEP tunnel.

\section{ACKNOWLEDGMENT}

We would like to thank R. Talman for many stimulating discussions and helpful remarks.


\begin{thebibliography}{9}   

\bibitem{LEP3} A. Blondel et al., ``LEP3: a High Luminosity $E^+E^-$ Collider to Study the Higgs Boson," CERN-ATS-NOTE-2012-061 TECH, August, 2012.

\bibitem{richter} B. Richter, ``Very high electron-positron colliding beams for the study of weak interactions,"
Nucl. Instr. Meth. {\bf 136} 47-60 (1976).

\bibitem{telnov} V. I. Telnov, ``Restriction on the energy and luminosity of $e^+e^-$ storage rings due to beamstrahlung,"
Phys. Rev. Lett. {\bf 110}, 114801 (2013).


\end{thebibliography}
\end{document}